\newif\ifanonymous \anonymousfalse
\newif\ifdraft \draftfalse
\newif\ifarxiv \arxivtrue
\newif\ifappendix \appendixtrue
\ifarxiv
    \ifanonymous
    \documentclass[acmtog,nonacm]{acmart}
    \else
    \documentclass[acmtog,nonacm]{acmart}
    \fi
\setcopyright{none}
\else
    \ifanonymous
    \documentclass[acmtog,anonymous,review]{acmart}
    \else
    \documentclass[acmtog]{acmart}
    \fi
    \acmSubmissionID{292}
\fi
\ifarxiv
\else
\setcopyright{cc}
\copyrightyear{2025}
\acmYear{2025}
\setcctype{by}
\acmConference[SIGGRAPH Conference Papers '25]{Special Interest Group on Computer Graphics and Interactive Techniques Conference Conference Papers }{August 10--14, 2025}{Vancouver, BC, Canada}
\acmBooktitle{Special Interest Group on Computer Graphics and Interactive Techniques Conference Conference Papers (SIGGRAPH Conference Papers '25), August 10--14, 2025, Vancouver, BC, Canada}
\acmDOI{10.1145/3721238.3730621}
\acmISBN{979-8-4007-1540-2/2025/08}
\fi

\usepackage{cleveref}
\usepackage{makecell}
\usepackage{soul}
\usepackage{subfigure}
\usepackage{xspace}
\usepackage{pifont} %
\crefname{section}{Sec.}{Secs.}
\Crefname{section}{Section}{Sections}
\Crefname{table}{Table}{Tables}
\crefname{table}{Tab.}{Tabs.}
\Crefname{figure}{Figure}{Figures}
\crefname{figure}{Fig.}{Figs.}

\ifdraft
\newcommand{\todo}[1]{{\color{red} #1}}

\newcommand{\src}[1]{\textcolor{violet}{SR: #1}}
\newcommand{\sr}[1]{{\color{violet} #1}}

\newcommand{\gtc}[1]{\textcolor{cyan}{GT: #1}}

\newcommand{\gtdel}[1]{{\color{cyan} \st{#1}}}
\newcommand{\dcc}[1]{{\color{red}\textbf{DC:} #1}}

\newcommand{\igcolor}{ForestGreen}
\newcommand{\igc}[1]{{\color{\igcolor}\textbf{IG:} #1}}

\newcommand{\yrc}[1]{{\color{orange}\textbf{YR:} #1}}

\newcommand{\abc}[1]{{\color{purple}\textbf{AB:} #1}}

\else
\newcommand{\todo}[1]{}
\newcommand{\src}[1]{}
\newcommand{\sr}[1]{#1}

\newcommand{\gtc}[1]{}

\newcommand{\gtdel}[1]{}

\newcommand{\dcc}[1]{}

\newcommand{\igc}[1]{}

\newcommand{\yrc}[1]{}

\newcommand{\abc}[1]{}

\fi

\newcommand{\algoname}{AnyTop\xspace}  %

\newcommand{\R}{\mathbb{R}} %
\newcommand{\N}{\mathbb{N}} %

\newcommand{\dd}{\mathcal{D}}

\newcommand{\mgg}{\mathcal{G}}
\newcommand{\mll}{\mathcal{L}}

\newcommand{\nn}{\mathcal{N}}
\newcommand{\pp}{\mathcal{P}}

\newcommand{\Rcal}{\mathcal{R}}
\newcommand{\mss}{\mathcal{S}}

\newcommand{\Crossgen}{Cross-skeleton generalization\xspace}
\newcommand{\crossgen}{cross-gen\xspace}
\newcommand{\Ingen}{In-skeleton Generalization\xspace}
\newcommand{\ingen}{in-gen\xspace}
\newcommand{\Stt}{Skeletal Temporal Transformer\xspace}
\newcommand{\stt}{STT\xspace}
\newcommand{\Unseengen}{Unseen-skeleton generalization\xspace}
\newcommand{\skelattn}{Skeletal Attention\xspace}
\newcommand{\tempattn}{Temporal Attention\xspace}
\makeatletter
\ifundef{\onedot} 
{
\DeclareRobustCommand\onedot{\futurelet\@let@token\@onedot}
\def\@onedot{\ifx\@let@token.\else.\null\fi\xspace}
}
{}
\ifundef{\eg} {\def\eg{\emph{e.g}\onedot}} {} \ifundef{\Eg} {\def\Eg{\emph{E.g}\onedot}} {} 
\ifundef{\ie} {\def\ie{\emph{i.e}\onedot}} {} \ifundef{\Ie} {\def\Ie{\emph{I.e}\onedot}} {} 
\ifundef{\cf} {\def\cf{\emph{cf}\onedot}} {} \ifundef{\Cf} {\def\Cf{\emph{Cf}\onedot}} {} 
\ifundef{\etc} {\def\etc{\emph{etc}\onedot}} {} \ifundef{\vs} {\def\vs{\emph{vs}\onedot}} {} 
\ifundef{\wrt} {\def\wrt{w.r.t\onedot}} {} \ifundef{\dof} {\def\dof{d.o.f\onedot}} {} 
\ifundef{\iid} {\def\iid{i.i.d\onedot}} {}  \ifundef{\wolog} {\def\wolog{w.l.o.g\onedot}} {} 
\ifundef{\etal} {\def\etal{\emph{et al}\onedot}} {} 
\makeatother

\newcommand{\cmark}{\ding{51}}
\newcommand{\xmark}{\ding{55}}
\ifappendix
  \newcommand{\supp}[1]{\Cref{#1}}
\else
  \newcommand{\supp}[1]{the sup. mat\onedot}
\fi

\begin{document}

\title{AnyTop: Character Animation Diffusion with Any Topology}

\author{Inbar Gat}
\authornote{Equal contribution.}
\affiliation{%
  \institution{Tel Aviv University}
  \country{Israel}
}
\email{gatinbar2344@gmail.com}
\author{Sigal Raab}
\authornotemark[1]
\affiliation{%
  \institution{Tel Aviv University}
  \country{Israel}
}
\author{Guy Tevet}
\affiliation{%
  \institution{Tel Aviv University}
  \country{Israel}
}
\author{Yuval Reshef}
\affiliation{%
  \institution{Tel Aviv University}
  \country{Israel}
}
\author{Amit H. Bermano}
\affiliation{%
  \institution{Tel Aviv University}
  \country{Israel}
}
\author{Daniel Cohen-Or} 
\affiliation{%
  \institution{Tel Aviv University}
  \country{Israel}
}
\renewcommand\shortauthors{Gat, I. et al.}
\begin{abstract}
Generating motion for arbitrary skeletons is a longstanding challenge in computer graphics, remaining largely unexplored due to the scarcity of diverse datasets and the irregular nature of the data. 
In this work, we introduce \algoname, a diffusion model that generates motions for diverse characters with distinct motion dynamics, using only their skeletal structure as input. 
Our work features a transformer-based denoising network, tailored for 
arbitrary skeleton
learning, integrating topology information into the traditional attention mechanism.
Additionally, by incorporating textual joint descriptions into the latent feature representation, \algoname learns semantic correspondences between joints across diverse skeletons. Our evaluation demonstrates that \algoname generalizes well, even with as few as three training examples per topology, and can produce motions for unseen skeletons as well. Furthermore, our model's latent space is highly informative, enabling downstream tasks such as joint correspondence, temporal segmentation, and motion editing.
Our webpage, \url{https://anytop2025.github.io/Anytop-page}, includes links to videos and code. %
\end{abstract}

\ccsdesc[500]{Computing methodologies~Motion processing}
\ccsdesc[300]{Computing methodologies~Computer graphics}
\ccsdesc[300]{Computing methodologies~Computer vision}
\ccsdesc[300]{Computing methodologies~Machine learning approaches}

\keywords{Animation, Motion synthesis, Deep Features, Computer Graphics.}
\begin{teaserfigure}
  \includegraphics[width=\textwidth]{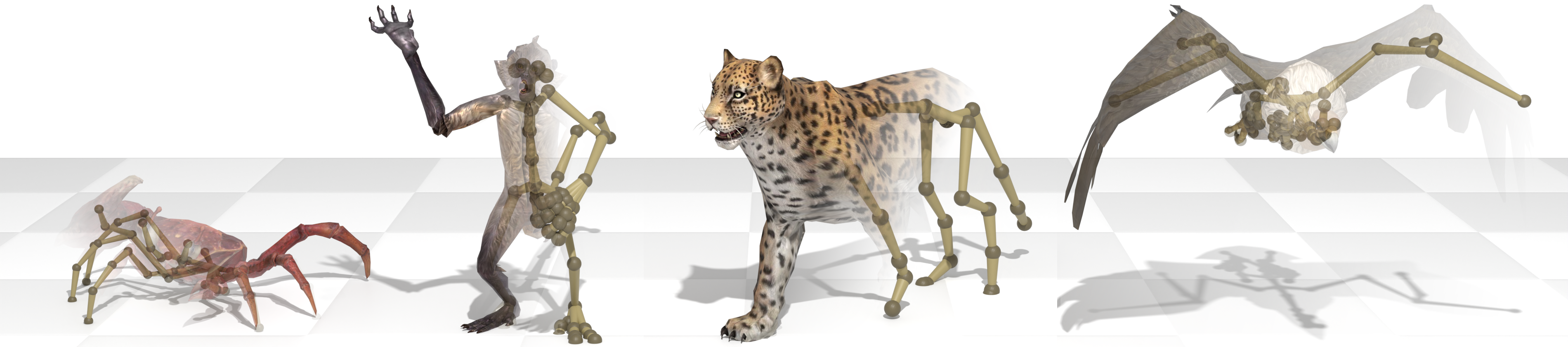}
  \caption{
\algoname generates motions for diverse characters with distinct motion dynamics, using only their skeletal structure \sr{and joint names} as input.
}
  \Description{}
  \label{fig:teaser}
\end{teaserfigure}

\maketitle
\section{Introduction}\label{sec:intro}
Character animation is a fundamental task in computer animation, playing a crucial role in industries such as film, gaming, and virtual reality. 
Animating 3D characters is a complex and time-consuming task that requires manual high-skill effort. 
Typically, animation pipelines involve a \emph{unique} skeleton for each character, defining its motion span, over which the animation is carefully crafted. 

In recent years, neural network-based approaches have simplified the animation process, showing impressive results in tasks such as motion generation and editing \cite{dabral2022mofusion,tevet2023human,holden2016deep,zhang2024motiondiffuse}. However, most existing methods cannot handle different skeletons and focus on a single topology \cite{shafir2024human,kapon2023mas}, target skeletons that differ only in bone proportions \cite{tripathi2024humos, yang2023omnimotiongpt}, or rely on skeletal homeomorphism \cite{aberman2020skeleton}. 

While effective within their scopes, these methods overlook the broader opportunities presented by diverse character animation, which require handling a wide variety of skeletal topologies. Conversely, methods designed to handle multiple skeletons often lack scalability, relying on topology-specific adjustments such as additive functional blocks for each skeleton \cite{li2024walkthedog} or entirely distinct instances of the model \cite{raab2024single, li2022ganimator}.

There are two main reasons keeping arbitrary skeleton animation generation largely under-explored. First, the irregular nature of the data, with skeletons varying in the number of joints and their connectivity, challenges standard methods for  processing and analysis. Second, the lack of datasets encompassing diverse topologies presents significant challenges for data-driven approaches.

In this work, we introduce \algoname, a diffusion framework designed to generate motions for arbitrary skeletal structures, as illustrated in \cref{fig:teaser}. \algoname is carefully designed to handle any skeleton in a general manner with no need for topology-specific adjustments.

\algoname is based on a transformer encoder, specifically adapted for graph learning. 
While many works embed an entire pose in one tensor \cite{han2024amd,xie2023omnicontrol}, we embed each joint independently at each frame \cite{aberman2020skeleton, agrawal2024skel}, enabling capturing both joint interactions within the skeleton and universal joint behaviors across diverse skeletal structures.
\algoname applies attention along both the temporal and skeletal axes. Notably, the skeletal attention is between \textit{all} joints. This is in contrast to previous art, and is made possible thanks to our topological conditioning scheme; 
we integrate graph characteristics \cite{park2022grpe, ying2021do}, such as joint parent-child relations, into the attention maps.
Consequently, each joint has access to information from all skeletal parts while also being able to prioritize topologically closer joints.
Furthermore, to bridge the gap between similarly behaved parts in different skeletons, \algoname incorporates textual descriptions of joints into the latent feature representation. 

\algoname is trained on Truebones Zoo dataset \cite{truebones}, which includes motion captures of diverse skeletal structures.
We contribute a processed version, aligned with the popular HumanML3D~\cite{guo2022generating} representation, which will be made publicly available. Using quantitative and qualitative evaluations, we show that \algoname outperforms current art.

\begin{table}[b]
    \caption{
    \textbf{Skeletal Variability.} 
    Character skeletons can vary in edge length, kinematic chain complexity, or overall topology. Each level of variation introduces greater challenges for motion synthesis. \algoname can generate motions for dozens of non-homeomorphic skeletons using a single model.
    }
    \centering
    \resizebox{\columnwidth}{!}{
    \begin{tabular} {l | c c c}
        \makecell[l]{Skeleton \\ Variability type}
        & \makecell{Edge lengths \\ variations} 
        & \makecell{ Kinematic chains \\ variations}
        & \makecell{Primal skeleton \\ variations} \\
        \midrule
        Single & \xmark& \xmark & \xmark  \\
        Isomorphic & \cmark& \xmark & \xmark   \\
        Homeomorphic & \cmark& \cmark & \xmark  \\
        Non-homeomorphic & \cmark& \cmark & \cmark  \\
        \bottomrule
    \end{tabular}
    } %
    \label{tab:skeletal_sim}
\end{table}

Our model demonstrates three forms of generalization in its generations: \emph{\Ingen} allows for new motion variants that preserve the character's original motion motifs; \emph{\Crossgen} 
facilitates generating motions that adapt motifs from several characters;
and \emph{\Unseengen} enables motion generation for skeletons not encountered during training.
Beyond its generative capabilities, \algoname's highly informative Diffusion Features (DIFT) \cite{tang2023emergent} 
enable
various downstream applications, including unsupervised correlation, temporal segmentation, and motion editing.

The approach presented here, and its ability to share information across characters, opens doors for 
more flexible generation, better equipped to learn and operate on more complex characters and scenarios, 
that better fit the real-world needs of 3D content creators.

\sr{\algoname makes the following key contributions: 
\begin{itemize}
    \item A new problem formulation for motion synthesis controlled by diverse skeletal structures.
    \item A novel architecture generalizing to non-homeomorphic skeletons with significant diversity.
    \item Comprehensive evaluation showing superior performance over baselines.
    \item  Careful analysis of learned features and generalization capabilities, establishing a foundation for interpretability and downstream tasks.
    \item A unified version of the Truebones zoo dataset with a preprocessing framework.
\end{itemize}
}

\section{Related Work}\label{sec:related_work}
\begin{figure*}

    \centering
    
    \includegraphics[width=\textwidth]{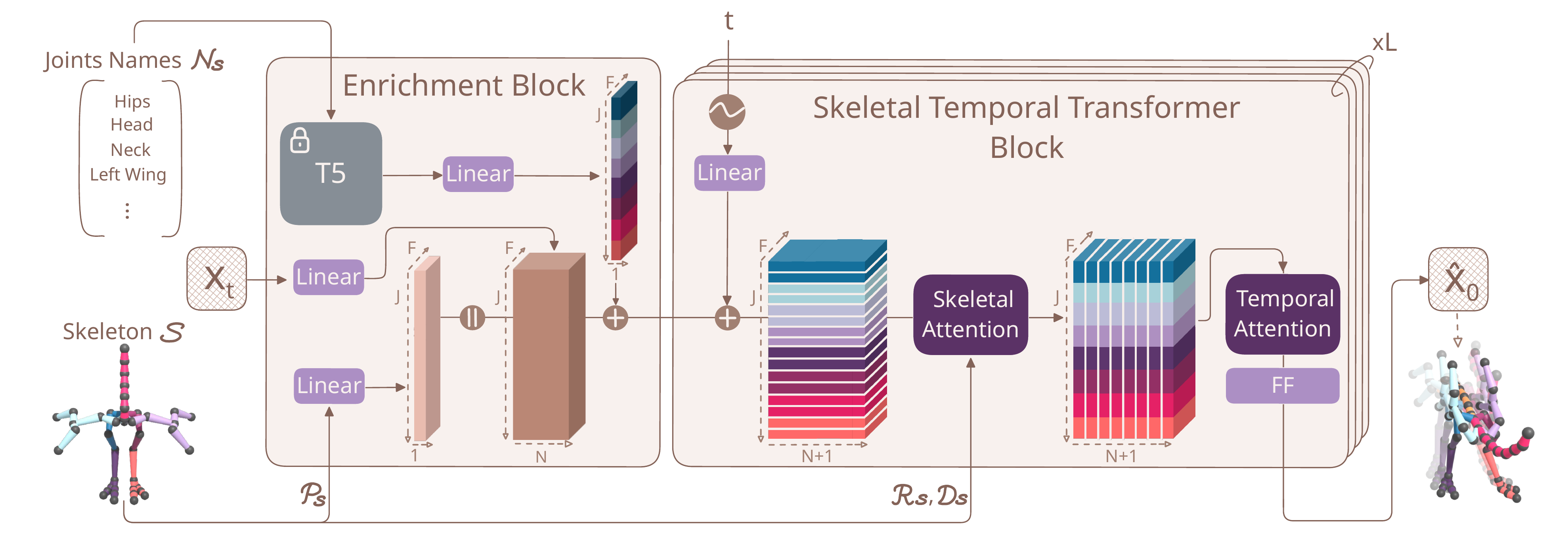}
    
    \caption{
        \textbf{Overview.}
        The input to \algoname is a noised motion $X_t$ and the skeleton $\mss=\{\pp_{\mss}, \Rcal_{\mss}, \dd_{\mss},\nn_{\mss}\}$, where $\pp_{\mss}$ refers to the rest-pose, $\Rcal_{\mss}$ denotes joints' relations, $\dd_{\mss}$ defines topological distances between each pair of joints and $\nn_{\mss}$ denotes joint names. The \textit{Enrichment Block} incorporates the skeletal features into the noised motion by concatenating the embedded $\pp_{\mss}$ to the sequence as an additional temporal token, and adding a T5-embedded name to each joint. The enriched motion is then passed through a stack of L \textit{\Stt} layers.  We apply skeletal attention along the joint axis to capture interactions between all joints, and incorporate topology information $\Rcal_{\mss}$ and $\dd_{\mss}$ to attention maps. Next, we apply temporal attention along the frame axis. Finally, the output is projected back to the motion features dimension, facilitating the reconstruction of the motion sequence.
    }
    \label{fig:arch}
    \Description[]{}  %
\end{figure*}

\paragraph{Skeletal variability in generative motion models }

We refer to four types of skeletal variability (\cref{tab:skeletal_sim}). The naming draws from terminology in the graph domain, hence we interchangeably use the terms joint and vertex, as well as edge and bone. 
A \emph{single} skeleton type refers to identical skeletons — that is, skeletons with the same vertices, connectivity, and edge lengths.
\emph{Isomorphic} skeletons correspond to isomorphic graphs, sharing vertices and edges but potentially differing in edge proportions.
\emph{Homeomorphic} skeletons may vary in structure, yet correspond to homeomorphic graphs, \ie,
use topologies obtained from the same primal graph by subdivision of edges. Specifically, homeomorphic skeletons share the same number of kinematic chains and end-effectors.
Finally, \emph{non-homeomorphic} skeletons vary in their structure and have no common primal graph. 

Most motion generative methods focus on a single skeletal structure \cite{raab2023modi, petrovich2021actor, karunratanakul2023gmd}. Others train on isomorphic skeletons \cite{zhang2023skinned,Villegas2021ContactAwareRO}, including works that use the SMPL \cite{loper2015smpl} body model
\cite{tripathi2024humos,jang2024geometry,petrovich2022temos} and SMAL \cite{zuffi20173d} 
body model
or its derivatives \cite{yang2023omnimotiongpt,rueegg2023barc}. A smaller portion of generative works support homeomorphic skeletons \cite{lee2023same,zhang2024skinned,cao2024car,studer2024factorized, ponton2024dragposer}. Among these works, some \cite{aberman2020skeleton} require a designated encoder and decoder per skeleton, and some \cite{zhang2024unified} 
offer a unified framework for all skeletons.

Only a handful of works can handle non-homeomorphic skeletons. \citet{martinelli2024moma} performs motion retargeting by learning a shared manifold for all skeletons, and decoding it to motions using learned skeleton-specific tokens.
The learned tokens capture the skeletal information of characters in the dataset, 
limiting the model's ability to generalize to skeletons unseen during training. Its results are shown exclusively on bipeds, leaving the applicability to other character families (\eg, quadrupeds, millipedes) unexplored.
WalkTheDog \cite{li2024walkthedog} uses a latent space that encodes motion phases and accommodates non-generative motion matching. 
\sr{A table clustering the methods mentioned above by supported skeletal variability can be found in \supp{sec:rw_supp}.}

A different class of generative models bypasses the handling of the skeletal structure by generating motion directly from point clouds \cite{mo2025motion}, shape-handles \cite{zhang2023tapmo} or meshes \cite{song2023unsupervised,ye2024skinned,muralikrishnan2025temporal,zhang2024dnf}. These works demonstrate great flexibility in target character structure, but overlook the advantage of skeletons, which are more compact and semantically meaningful, easier to manipulate via rig-based animation, and compatible with physics engines \cite{tevet2024closd} and inverse kinematics systems. Some works \cite{wang2024mmr} perform automatic rigging after the generation, but automatic rigging often necessitates manual adjustments.

Finally, methods that support arbitrary skeletons \cite{raab2024single,li2022ganimator} involve a separate training process for each skeleton, exhibiting scaling issues and lacking \Crossgen.

\algoname addresses training on non-homeomorphic skeletons and is the only skeletal-based approach capable of generating natural, smooth motions on a diverse range of characters, including bipeds (e.g., raptor, bird), quadrupeds (e.g., dog, bear), multi-legged arthropods (e.g., spider, centipede), and limbless creatures (e.g., snakes). To the best of our knowledge, our work is the only one capable of accepting an input topology, including unseen ones, and generating motions based on that topology.

\paragraph{Transformer-based Graph Learning}
Early versions of deep networks on graphs relied on convolutional architectures \cite{kipf2016semi}. The emergence of transformers has sparked a new avenue of research, integrating graphs and transformers.
GAT \cite{velivckovic2017graph}
replace the graph-convolution operation with a self-attention module, where attention is restricted to neighboring nodes. \citet{rong2020self} iteratively stack self-attention layers alongside graph convolutional ones to account for long-range interactions between nodes.
Unlike transformers in the language and imaging domains, and due to the irregular structure of graphs, these earlier works do not use positional encoding.

Subsequent works \cite{dwivedi2021generalization,kreuzer2021rethinking} linearize the graphs into an array of nodes and add absolute positional encoding to each node.
However, linearization is unnatural to the graph structure, requiring a reconsideration of the approach.

Encoding relative positional information has been explored to maintain positional precision while adhering to the graph's structure. Works using it \cite{ying2021transformers, shaw2018self, park2022grpe} integrate relative positional encoding into the attention map based on relative measures, such as shortest path distance between nodes or edge type.

The aforementioned approaches are \emph{discriminative}, applied to tasks such as regression and segmentation.
\algoname leverages the relative positional encoding approach for \emph{generative} tasks and tailors it to the \emph{motion}
domain. In particular, our work redefines edge types to capture joint relations within skeletal structures and considers a temporal axis, which is not present in the graph domain.

\section{Method}\label{sec:method}

\algoname is a diffusion model synthesizing motions for multiple different characters with arbitrary skeletons. Given a skeletal structure for input, it generates a natural motion sequence with high fidelity to ground-truth characters.
\algoname is based on a transformer encoder, specifically adapted for graph learning, as depicted in \cref{fig:arch}.

\subsection{Preliminaries} \label{sec:prelim}
\ifarxiv
\else
\begin{figure}
    \centering
    
\includegraphics[width=\columnwidth]{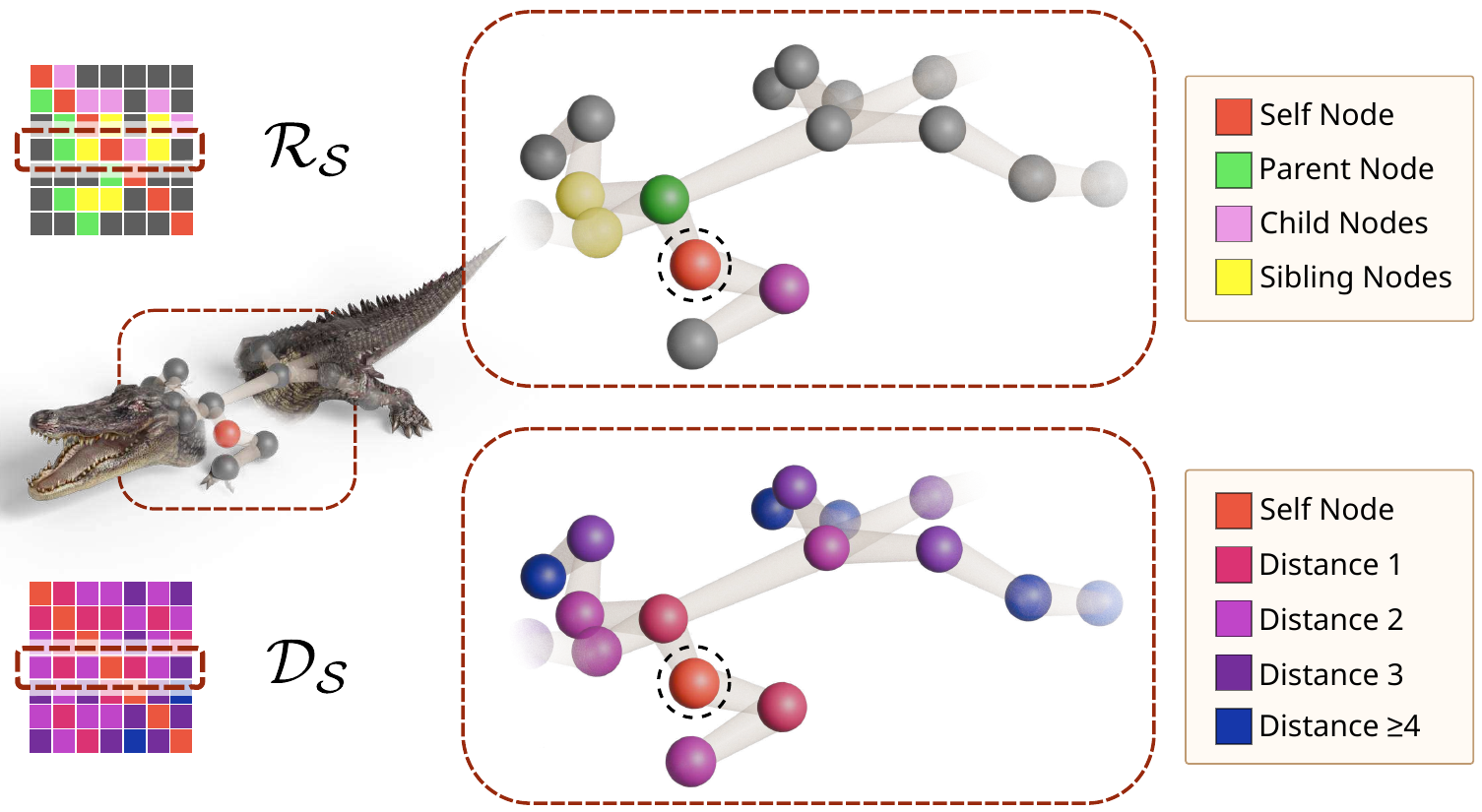}
    \caption{
        \textbf{Topological Conditions.}
        Joint relations $\Rcal_{\mss}$ (top) and graph distances  $\dd_{\mss}$ (bottom), visualized for a specific joint marked in red. Different colors indicate different values in the row corresponding to the visualized joint in the $\Rcal_{\mss}, \dd_{\mss}$ matrices.
       }
    \label{fig:topo_cond}
    \Description[]{}  %
\end{figure}

\fi
\paragraph{Motion Representation}
We represent motion as a 3D tensor $X\in \R ^ {N \times J \times D}$, where $N$ and $J$ are the maximum number of frames and joints across all motions in the dataset, and D is the number of motion features per joint. 
As motions vary in duration and skeletal structure, we pad the original number of frames and joints of each motion to match the maximum values $N$ and $J$, respectively.
We adopt a redundant representation, where each joint $j$ (except the root) consists of its root-relative position $p_j\in\R^3$, 6D joint rotation $r_j\in\R^6$ \cite{DBLP:journals/corr/abs-1812-07035}, linear velocity $v_j\in\R^3$, and foot contact label $fc_j \in \{0, 1\}$.
Altogether a joint is represented by $\{p_j,r_j,v_j,fc_j\}\in\R^{13}$, hence
 $D=13$. 
Our representation is inspired by \citet{guo2022generating}; however, our approach maintains features at the joint level by representing each joint as a separate tensor, yielding $J$ tokens per frame. Conversely, \citeauthor{guo2022generating} concatenate features from all joints into one tensor, producing one token per frame.

\paragraph{Skeletal structure Representation}
In the context of 3D motion, \emph{topology} is a directed, acyclic, and connected graph (DAG). Adding geometric information to this graph makes it a \emph{skeleton}. 
We use the terms ``topology" and ``skeleton" interchangeably throughout this work, clarifying any distinction when necessary.
A \emph{rest-pose} is the character's natural pose, represented by $(\mgg, O)$, where $\mgg$ is a DAG defining the topological hierarchy and $O\in \R^{J\times 3}$ is a set of 3D offsets, specifying each joint's parent-relative position.
In our work, we represent a skeleton by $\mss = \{\pp_\mss, \Rcal_\mss, \dd_\mss, \nn_\mss\}$. The first term, $\pp_\mss \in \R^{J \times D}$, is the rest-pose, converted to the format of individual poses in the motion sequence. The second term, $\Rcal_\mss\in \N_0^{J\times J}$, is the joints relations, where $\Rcal_\mss[i, j]$ holds the relation type between $i$ and $j$. We allow six types of relations, which are \emph{child}, \emph{parent}, \emph{sibling}, \emph{no-relation}, \emph{self} and \emph{end-effector}. \emph{Self} and \emph{end-effector} are valid only in case $i=j$, and \emph{end-effector} specifies if the joint is a leaf in $\mgg_\mss$. 
The third term, $\dd_\mss\in \N_0^{J\times J}$, represents the graph distances, where $\dd_\mss[i,j]$ holds the topological distance between $i$ and $j$ in $\mgg_\mss$, up to a maximal distance $d_{max}$. 
The topological conditions, $\Rcal_\mss$ and $\dd_\mss$, are illustrated in \cref{fig:topo_cond}.
Finally, $\nn_\mss$ is the joints' textual descriptions, which are typically included in 3D asset formats (\eg, bvh, fbx). 
\ifarxiv
\begin{figure}
    \centering
    
\includegraphics[width=\columnwidth]{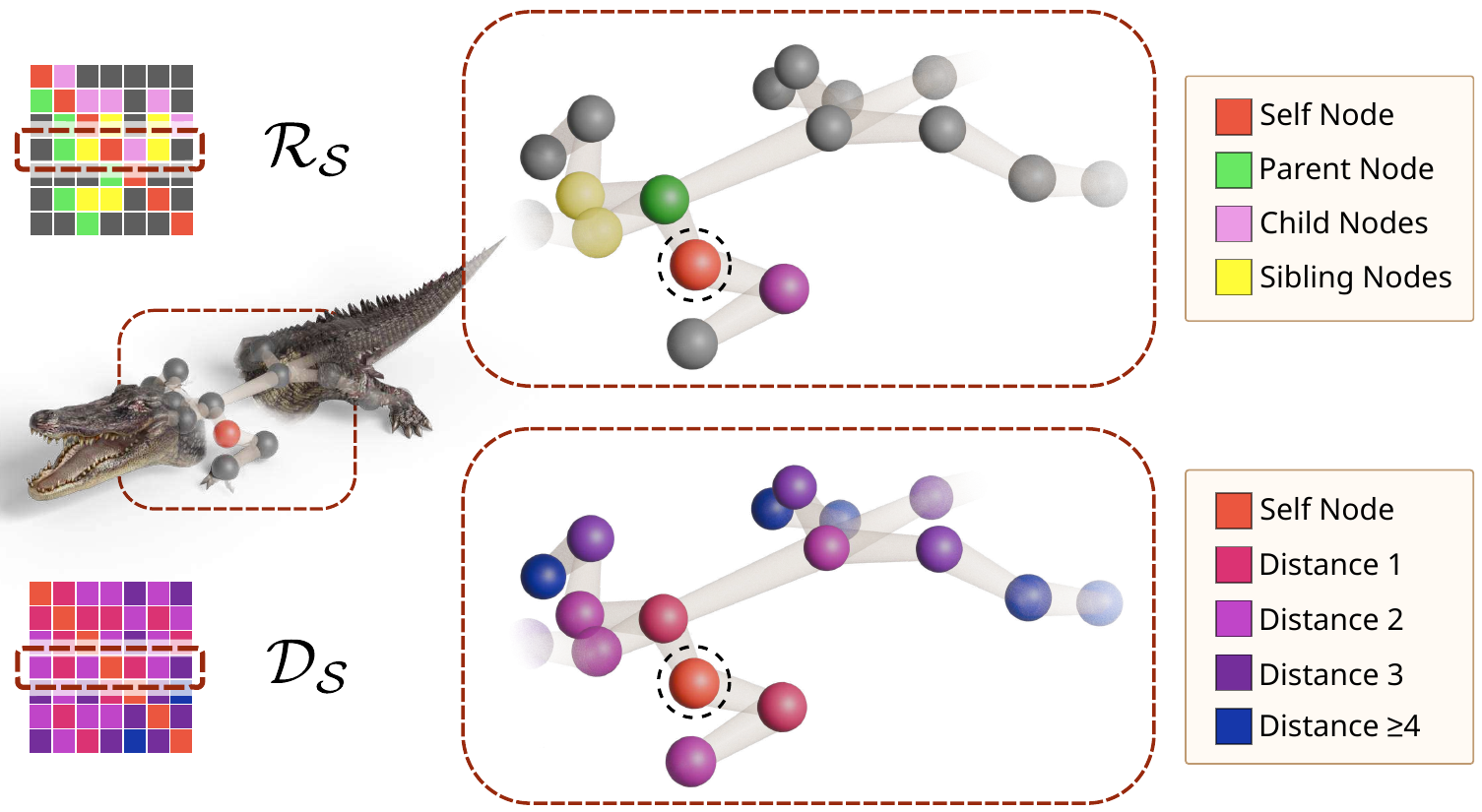}
    \caption{
        \textbf{Topological Conditions.}
        Joint relations $\Rcal_{\mss}$ (top) and graph distances  $\dd_{\mss}$ (bottom), visualized for a specific joint marked in red. Different colors indicate different values in the row corresponding to the visualized joint in the $\Rcal_{\mss}, \dd_{\mss}$ matrices.
       }
    \label{fig:topo_cond}
    \Description[]{}  %
\end{figure}

\else
\fi
\subsection{Architecture}

\algoname is a generative Denoising Diffusion Probabilistic Model (DDPM) \cite{ho2020denoising}.
At each denoising step $t \in [1,T]$ it gets a noisy motion $X_t$ and a skeleton $\mss=\{\pp_\mss, \Rcal_\mss, \dd_\mss, \nn_\mss\}$ as input, and predicts the clean motion $\hat{X}_{0}$ \cite{tevet2023human} rather than the noise $\epsilon_t$.

\algoname consists of two primary components, illustrated in \cref{fig:arch}. The first is an \emph{Enrichment Block}, which integrates skeleton-specific information into the noised motion. The second is a \emph{\Stt Block}, which employs attention across both skeletal and temporal axes while embedding topological information into the skeletal attention maps.

\paragraph{Enrichment block}
This block incorporates semantic information from the rest-pose $\pp_\mss$ and the joint descriptions $\nn_\mss$, into the noised sample $X_t$. 
It projects $\pp_\mss$ to \sr{latent feature size} $F$ and concatenates it with $X_t$ along the temporal axis, effectively making it frame 0. The joint descriptions $\nn_\mss$ are encoded by a T5 model \sr{\cite{JMLR:v21:20-074}}, projected to length $F$, and added to their corresponding joint features across all frames.
Finally, the block outputs the enhanced data of shape $\R^{(N+1) \times J \times F}$.
 
\paragraph{\Stt block}
The inputs to this block are the embedded tokens of $X_t$ emitted from the \emph{Enrichment Block}, the diffusion step $t$, and the precomputed values $\dd_\mss$, $\Rcal_\mss$. 
The block comprises a stack of $L$ identical \emph{Skeletal Temporal Transformer} (STT) encoder layers, each consisting of three parts.
The first component is a \emph{\skelattn}, which performs spatial self-attention across joints within the same frame. Unlike concurrent approaches that limit attention or convolution to adjacent joints within the skeletal hierarchy, our method enables each joint to attend to all others, capturing long-range relations. To regain the local joint knowledge, we incorporate topology information $\Rcal_\mss$ and $\dd_\mss$ into the attention maps. This allows each joint to access information from all skeletal parts while also prioritizing topologically closer joints.

The second component is a \emph{\tempattn}, which applies self-attention along the temporal axis for each joint independently, observing its motion over time. 
To enhance efficiency and mitigate overfitting, the temporal attention is applied within a temporal window of length $W$. 
The third component is a feed-forward block.
Finally, the output is projected to the original motion dimension, enabling motion reconstruction.
\paragraph{Topological Conditioning Scheme}
We extend transformers for graph-based learning by incorporating both graph topology and node interaction information through our \emph{Skeletal Attention} mechanism. Inspired by \emph{discriminative} works in the \emph{graphs} domain \cite{ying2021transformers}, \algoname introduces a novel method for \emph{generative} tasks, specifically tailored to the \emph{motion} domain.
We integrate graph properties directly into attention maps, enabling the structural characteristics of the graph to influence the learning process. 
Our work uses two types of node affinity, the topological distance, $\dd_\mss$, and relations, $\Rcal_\mss$, as detailed in \cref{sec:prelim}.
We incorporate the graph information into the attention maps \cite{park2022grpe}, by learning distinct query and key embeddings for distances, denoted by 
$E^\dd_q$, $E^\dd_k \in \R^{(d_{max}+1) \times F}$,
and embeddings for relation, denoted by 
$E^\Rcal_q$, $E^\Rcal_k \in \R^{6\times F}$, 
where $E^{(\cdot)}_q$ and $E^{(\cdot)}_k$ denote embeddings that relate to queries and keys, respectively, and $F$ is the latent feature size. 
These embeddings are used to form two new attention maps, $a^{\dd}$ and $a^{\Rcal}$ defined for a given pair of joints $i, j \in [J]$: 
    \begin{align}
        a^{\dd}_{ij}=q_i \cdot E^\dd_q [\dd_{ij}] + k_j \cdot E^\dd_k [\dd_{ij}],
    \end{align}
    \begin{align}
        a^{\Rcal}_{ij}=q_i \cdot E^\Rcal_q [\Rcal_{ij}] + k_j \cdot E^\Rcal_k [\Rcal_{ij}],
    \end{align}
where $q_i$, $k_j$ denote the $i$'th joint query and $j$'th joint key, respectively, and $[\cdot]$ denotes an index in the embedding matrix.
Finally, we incorporate graph information by adding the two attention maps to the standard attention map and scaling their sum:
\begin{align}
        a_{ij} = \frac{q_i \cdot k_j + a^{\dd}_{ij} + a^{\Rcal}_{ij}}{\sqrt{F}}.
\end{align}
The final attention score is computed by applying the standard row-wise softmax to $a_{ij}$.
\subsection{Training}\label{subsec:training}
\paragraph{Data Sampling and Augmentations}
We train \algoname using minibatches sampled with a \emph{Balancing Sampler} to mitigate data imbalance and the dominance of specific skeletons. 
\sr{Given $k$ skeletal types, with $n_i$ samples corresponding to skeletal type $i$, our balancing sampler assigns a sampling probability of $1/(n_i\cdot k)$ to each instance of type $i$.}
To further enhance generalization, we apply skeletal augmentations, including randomly removing 10\% to 30\% of the joints and adding new joints at the midpoint of existing edges. More details on data augmentation are provided in \supp{sec:data}.

\paragraph{Training Objectives}
Given a motion $X_0$ of skeleton $\mss$, its noised counterpart $X_t$, with diffusion step $t\sim [1,T]$, our model predicts the clean  motion,  $\hat{X}_{0} = \algoname(X_{t}, t, \mss) $.
Our main objective is defined by the \emph{simple} formulation \cite{ho2020denoising}, namely,
\begin{align}
   \mll_{simple} =  E_{t \sim [1,T]}\left\lVert \sr{\algoname(X_{t}, t, \mss)} - X_{0}\right\rVert^2_2.
\end{align}

The Mean Squared Error (MSE) over rotations does not directly correlate to their distance in the rotation space, hence we apply a geodesic loss \cite{Huang_2017_CVPR, tripathi2024humos} over the learned rotations. Let $r, \hat{r} \in \R^{N\times J \times 6}$ denote the 6D rotations of $X_0$ and $\hat{X}_0$ respectively. The geodesic loss is defined as follows: 
\begin{align}
    \mathcal{L}_{rot} = \sum_{n=1}^N {\sum_{j=1}^J {\arccos\frac{Tr(GS(r_{n,j})(GS(\hat{r}_{n,j})^T) - 1 }{2}}},
\end{align}
where $GS$ is the Gram-Schmidt process, used to convert 6D rotations to rotation matrices \cite{zhou2019continuity}, and $Tr$ is the matrix Trace operation. 
Overall, the final training objective is 
\begin{align}
    \mathcal{L} = \mll_{simple} + \lambda_{rot}\mll_{rot}.
\end{align}

\section{Analysis}\label{sec:analysis}

\subsection{Latent Space Analysis}
In this section, we examine \algoname's latent space and show that it features a unified manifold for joints across all skeletons.
We use DIFT \cite{tang2023emergent}, a framework designed for detecting correspondence in the latent space of models undergoing diffusion.
DIFT features are intermediate activations from layer $l_{corr}$, extracted during a single denoising pass on a sample that has been noised directly to diffusion step $t_{corr}$. These features serve as effective semantic descriptors for predicting correspondence.
Note that the values we choose for $l_{corr}$ and $t_{corr}$ align with those used in the original DIFT work.
Let $X^{ref}$ denote a reference motion, and let $X^{tgt}$ denote a motion in which we search for corresponding parts.
Let $S^{ref}, S^{tgt}$ denote their skeletons, respectively.

Our spatial and temporal correspondence results are illustrated in \cref{fig:spatial_cor,fig:temoral_cor}, respectively, and in the supplementary video.

\paragraph{Spatial Correspondence}
\begin{figure}
    \centering
    \includegraphics[width=0.8\columnwidth]{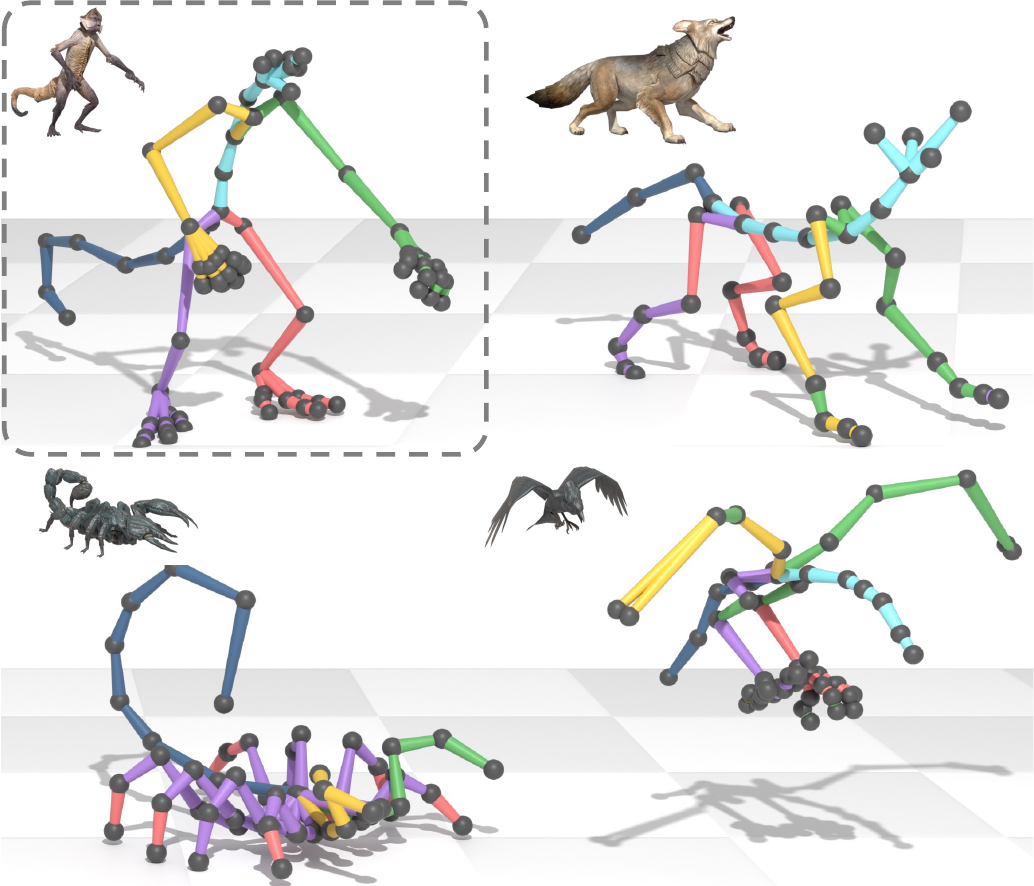}
    
    \caption{
        \textbf{Spatial Correspondence.} Monkey (top left) depicts the reference skeleton, while the fox, scorpion, and bird depict different target skeletons. 
        Target skeleton joints are color-coded to match their corresponding joints in the reference. 
        For better visualization, we color the bones to match their adjacent joints. 
        Note the correspondence in limbs, spine, and tail.
       }
    \label{fig:spatial_cor}
    \Description[]{}  %
\end{figure}

\begin{figure}

    \centering
    
    \includegraphics[width=\columnwidth]{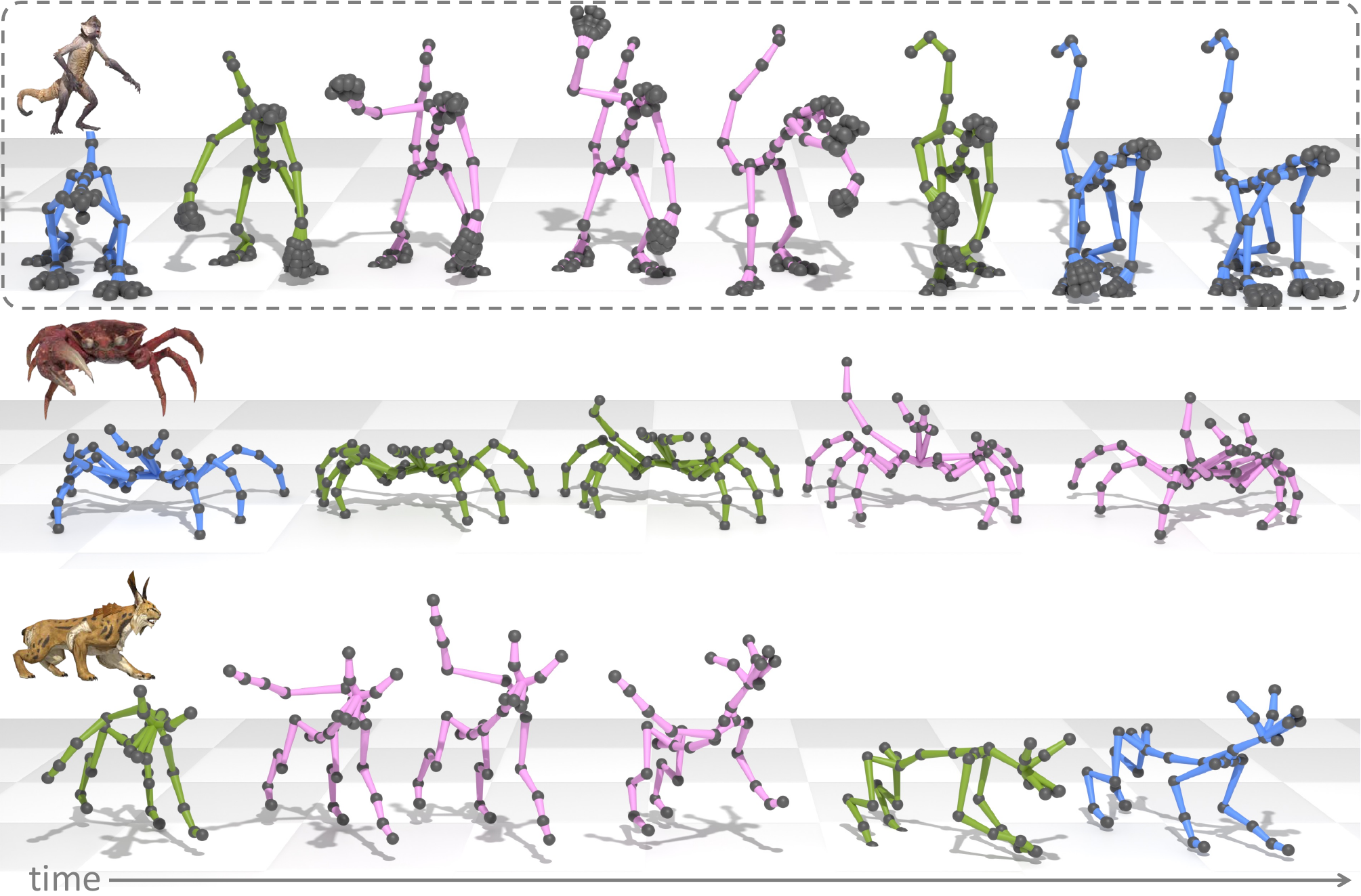}
    
    \caption{
    \textbf{Temporal Correspondence.} Monkey (top row) features the reference motion, while the Crab and Lynx represent two target motions. The frames of the targets are color-coded to align with their corresponding reference frames.
    Note the correspondence: aggressive motion segments are pink, idle frames blue, and transitional frames green.
    }
    \label{fig:temoral_cor}
    \Description[]{}  %
\end{figure}

We show that manifold features of semantically similar skeletal joints across different characters are close to each other (\cref{fig:spatial_cor}).
Our objective is to find the most similar \emph{joint} in $\mss^{ref}$ for each joint in $\mss^{tgt}$.
To achieve this, we extract DIFT features for both motions $X^{ref}, X^{tgt}$ at diffusion step $t_{corr} = 2$ and layer $l_{corr} = 0$, average them along the temporal axis, and obtain a single feature vector per joint.
Using cosine similarity, we detect the closest counterpart for each joint in $S_{tgt}$. 

\sr{
\cref{fig:spatial_cor} shows a clear alignment between corresponding body parts (\eg, limbs, spine, tail) across different animals, with only minor confusion related to left-right distinction.
Left-right issues are common in feature extraction \cite{zhang2024telling}, to be addressed directly.
}

\paragraph{Temporal Correspondence}
We show that \algoname can recognize pose-level similarities and identify analogous actions across different skeletons (\cref{fig:temoral_cor}).
This time, our objective is to find the most similar \emph{frame} in $X^{ref}$ for each frame in $X^{tgt}$.
To accomplish this goal, we extract DIFT features at diffusion step $t_{corr}=3$ and layer $l_{corr}=1$, and average them along the skeletal axis, resulting in a single feature vector per frame.
We use cosine similarity to detect the closest counterpart for each frame in $X_{tgt}$.

\begin{figure}

    \centering
    
    \includegraphics[width=\columnwidth]{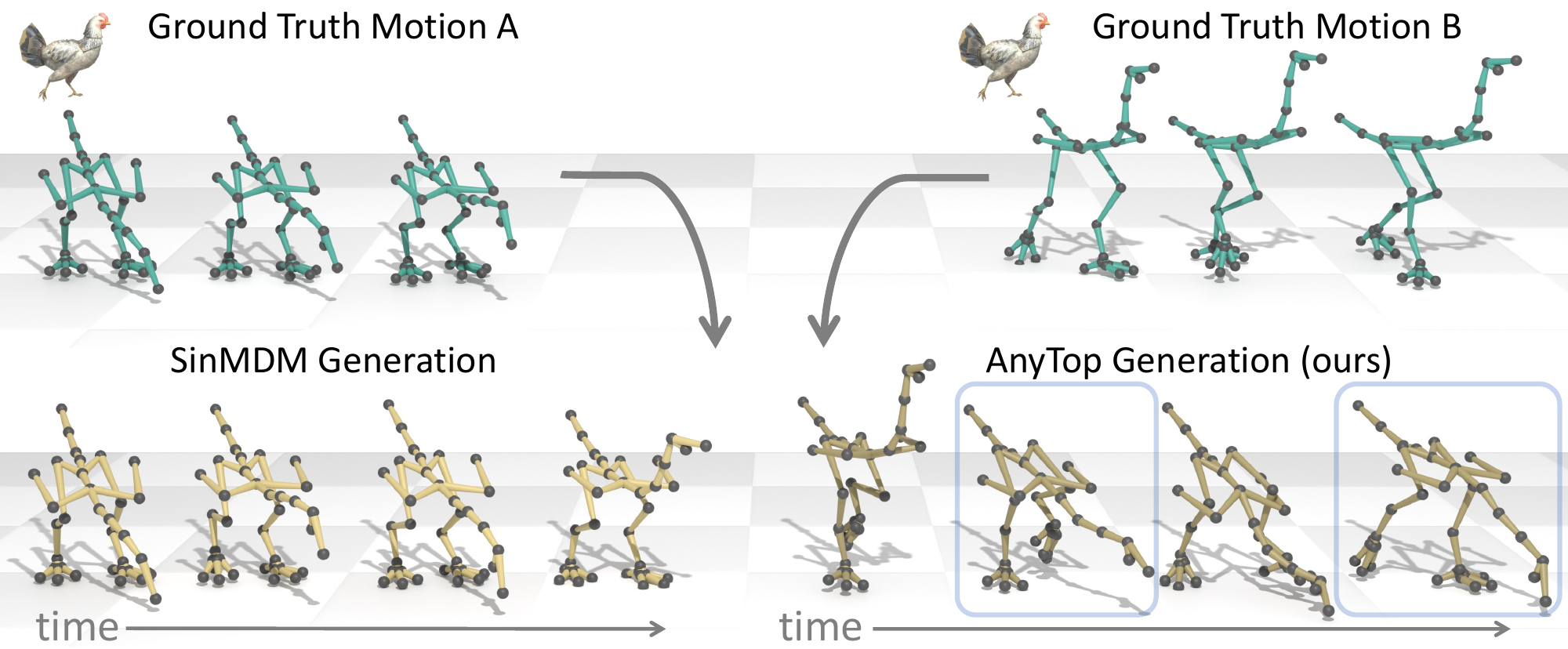}
    
    \caption{
    \textbf{\Ingen.} The top row depicts two ground truth chicken motions: pecking (left) and walking (right). The bottom row presents synthesized motions of an adapted SinMDM (left) and \algoname (right). The emphasized frames in \algoname demonstrate spatial composition of walking and pecking, introducing novel poses not present in the ground truth. SinMDM embeds entire poses, hence cannot spatially-compose joints. 
    }
    \label{fig:in_gen}
    \Description[]{}  %
\end{figure}

\begin{figure}

    \centering
    
    \includegraphics[width=\columnwidth]{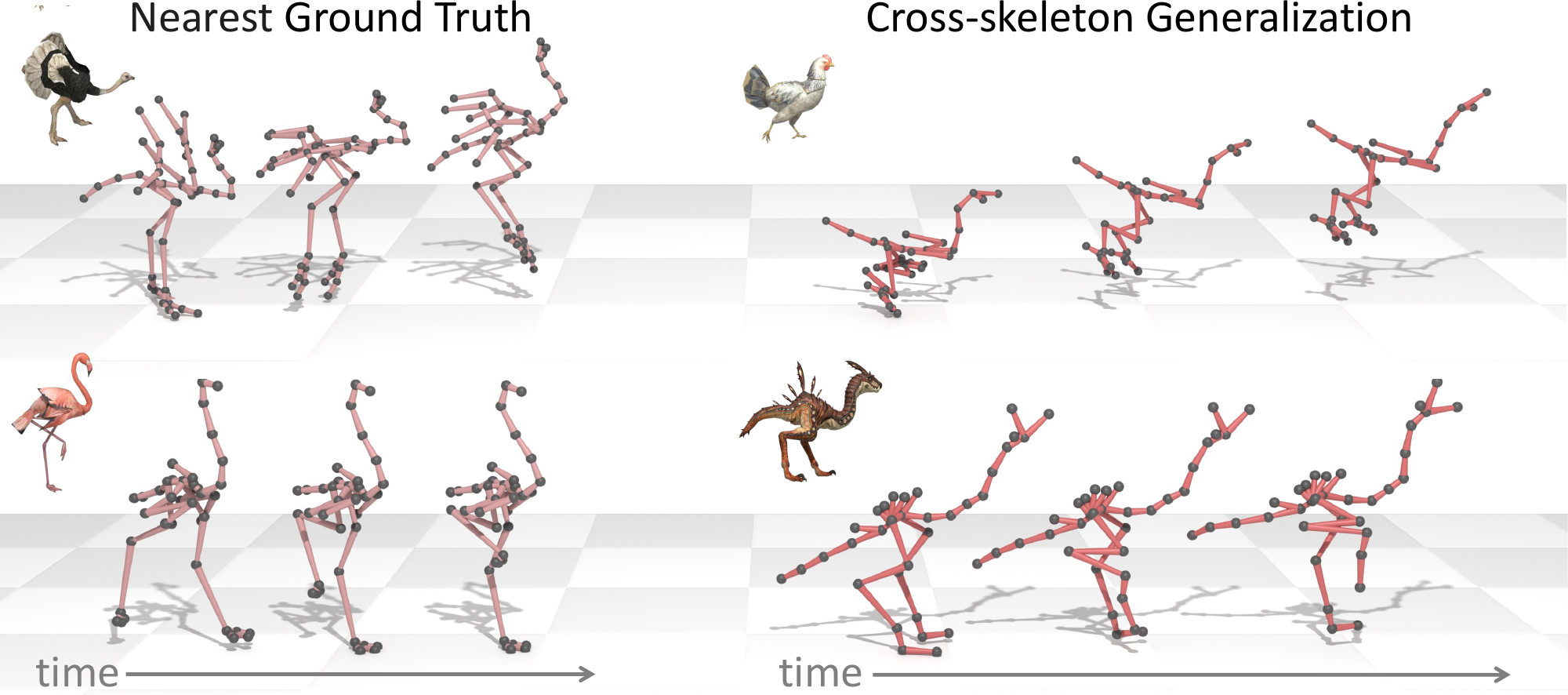}
    
    \caption{
        \textbf{\Crossgen.} 
        Right: a generated motion featuring an action not in the performing skeleton’s ground truth.
        Left: notably, the nearest ground truth originates from a different character.
    }
    \label{fig:cross_gen}
    \Description[]{}  %
\end{figure}

\subsection{Generalization Forms}

We identify three forms of generalization in our generated motions.

\paragraph{\Ingen} dubbed \emph{\ingen}, refers to generalization within a specific skeleton, featured as both \emph{temporal composition} -- combining motion segments from dataset instances, and \emph{spatial composition} -- introducing novel poses by combining skeletal parts of ground truth poses. Notably, \emph{spatial composition} is enabled by our per-joint encoding, which provides the flexibility required for such diversity. In \cref{fig:in_gen} and in our supp. video, we showcase \algoname's \ingen and highlight how other methods, which embed the entire pose, fail to achieve a comparable variety.

\paragraph{\Crossgen} dubbed \emph{\crossgen}, 
is expressed through shared motion motifs across different characters. This form of generalization enables the adaptation of motion behaviors originally performed by other skeletons, as shown in \cref{fig:cross_gen} and our supp. video. 
When motions must strictly align with typical behaviors, the training dataset can be restricted accordingly.

\begin{figure}

    \centering
    
    \includegraphics[width=\columnwidth]{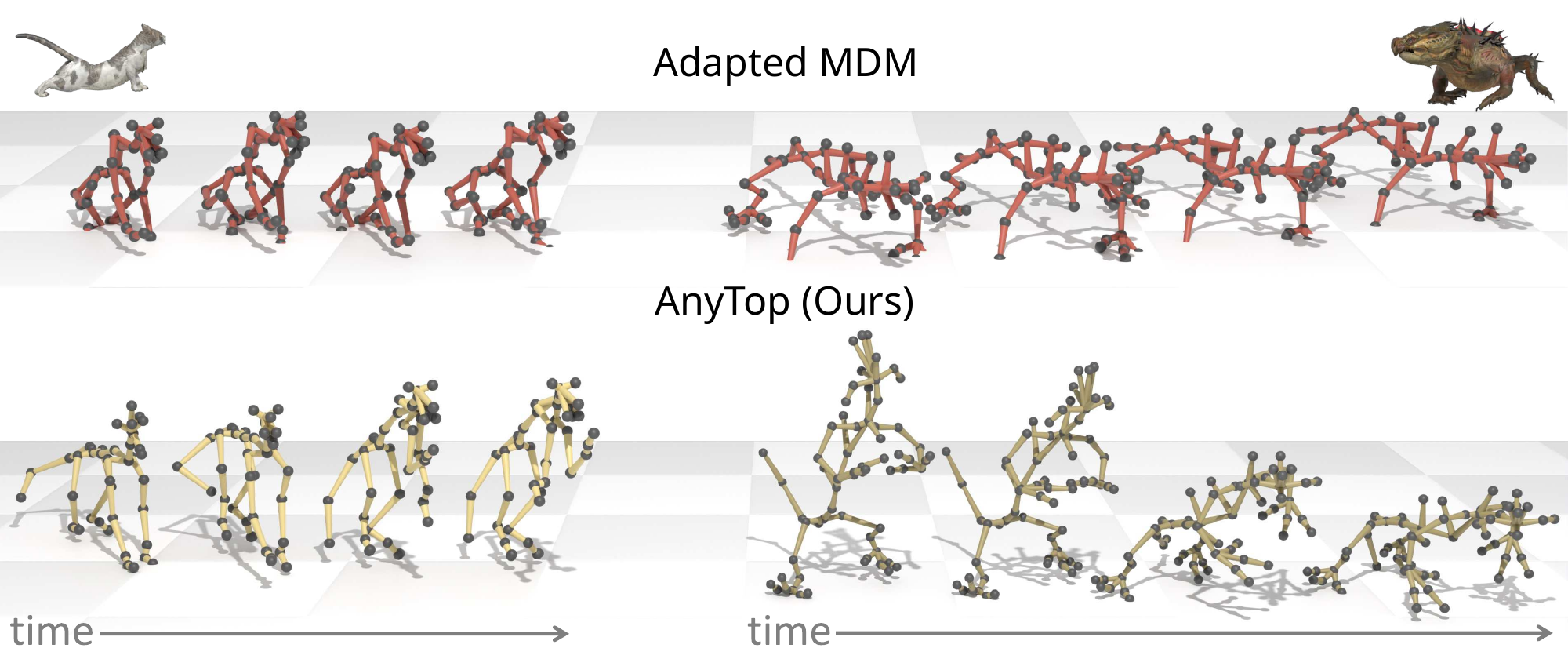}
    
    \caption{
        \textbf{\Unseengen} 
        Zero-shot inference of the cat (left) and komodo dragon (right) using \sr{adapted MDM baseline (top) and \algoname (bottom)}. \algoname's generated motions maintain natural appearance while MDM's generated motions are static and jittery. 
        }
    \label{fig:unseen}
    \Description[]{}  %
\end{figure}

\paragraph{\Unseengen} 
extends to skeletons not encountered during training and illustrated in \cref{fig:unseen} and the video.
\sr{We assess generalization to varying levels of skeletal out-of-distribution (OOD) using three insect characters: Crab (in-distribution), Centipede (moderately OOD), and Cobra (highly OOD). We quantify the OOD degree using Wasserstein distance on graph features. Each character is evaluated in both seen and unseen settings.
Results, in \cref{tab:unseen_ood}, demonstrate robust performance on unseen skeletons that are in-distribution, with a gradual decline as the skeletons deviate from the training distribution. 
The decline is driven by divergence from the data distribution, featured by a drop in fidelity (coverage, intra-diversity difference) and a rise in local/inter-diversity.}

\begin{table}[b]
    \caption{
    \sr{
    \textbf{Unseen-skeleton generalization vs. OOD degree.} 
    Performance on unseen skeletons gradually drops as they deviate from data distribution. 
    }
    }
    \centering
    \resizebox{\columnwidth}{!}{
    \begin{tabular} {l | c c c c c}
         Character & \makecell{OOD\\ Score} & 
         \makecell{Coverage $\uparrow$\\ Seen / Unseen}  & \makecell{Local Div. $\uparrow$\\ Seen / Unseen} & \makecell{Inter Div. $\uparrow$\\ Seen / Unseen} & \makecell{Intra Div. Diff.$\downarrow$\\ Seen / Unseen} \\
        \midrule
       Crab & 1.70 & 99.45 / 88.69 & 0.22 / 0.29 & 0.34 / 0.36 & 0.13 / 0.14 \\   
       Centipede & 2.26 & 83.55 / 43.90 & 0.28 / 0.29 & 0.41 / 0.16 & 0.12 / 0.15 \\
       Cobra & 4.04 & 79.34 / 16.47 & 0.17 / 0.56 & 0.39 / 1.03 & 0.20 / 0.46 \\
        \bottomrule
    \end{tabular}
    } %
    \label{tab:unseen_ood}
\end{table}

\section{Applications} \label{sec:applications}
\algoname enables various downstream tasks; we demonstrate two.

\begin{figure}

    \centering
    
    \includegraphics[width=\columnwidth]{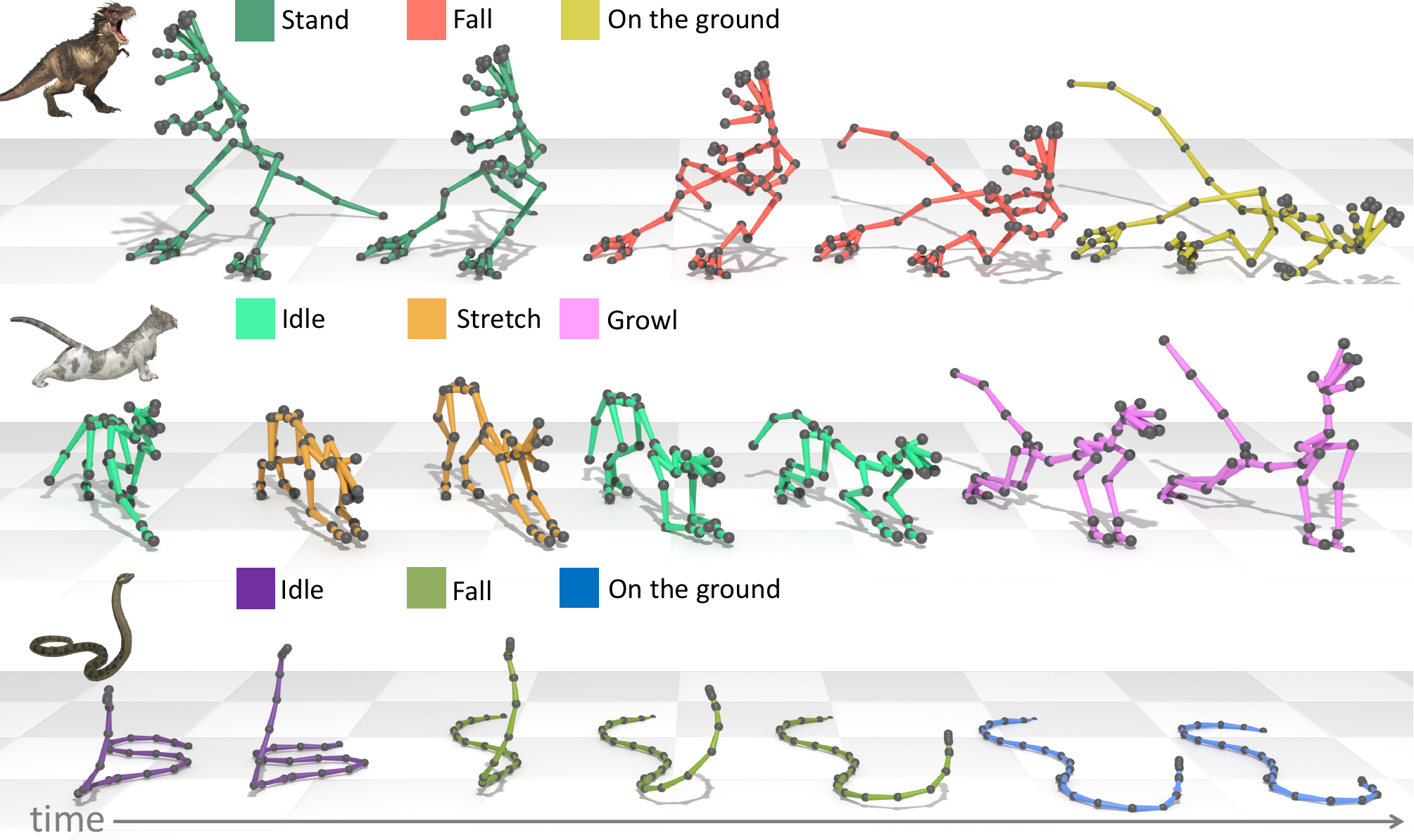}
    
    \caption{
        \textbf{Temporal Segmentation.} 
        Temporal clustering on a tyrannosaurus, a cat, and an anaconda snake, using K-means on PCA-reduced DIFT features.
    }
    \label{fig:temoral_seg}
    \Description[]{}  %
\end{figure}

\paragraph{Temporal Segmentation}Temporal segmentation is the task of partitioning a temporal sequence into disjoint groups, where frames sharing similar characteristics are grouped. 
For a clean sample $X_{0}$, either generated or given, and skeleton $\mss$, we extract DIFT features at diffusion step $t_{seg}\!=\!\!3$ and layer $l_{seg}\!=\!\!1$. The features are averaged along the joint dimension to produce a single feature vector per frame. We apply PCA for dimensionality reduction and then use K-means to cluster the frames into $k\!\!=\!\!3$ categories. Our results are visualized in \cref{fig:temoral_seg} and in the supp. video.
This application reinforces \cref{sec:analysis}, showing that \algoname's latent features are effective frame descriptors.
However, in \cref{sec:analysis}, frames are grouped by similarity to $X^{ref}$, while here they are grouped by similarity to each other.

\paragraph{Editing} 

\begin{figure}

    \centering
    
    \includegraphics[width=\columnwidth]{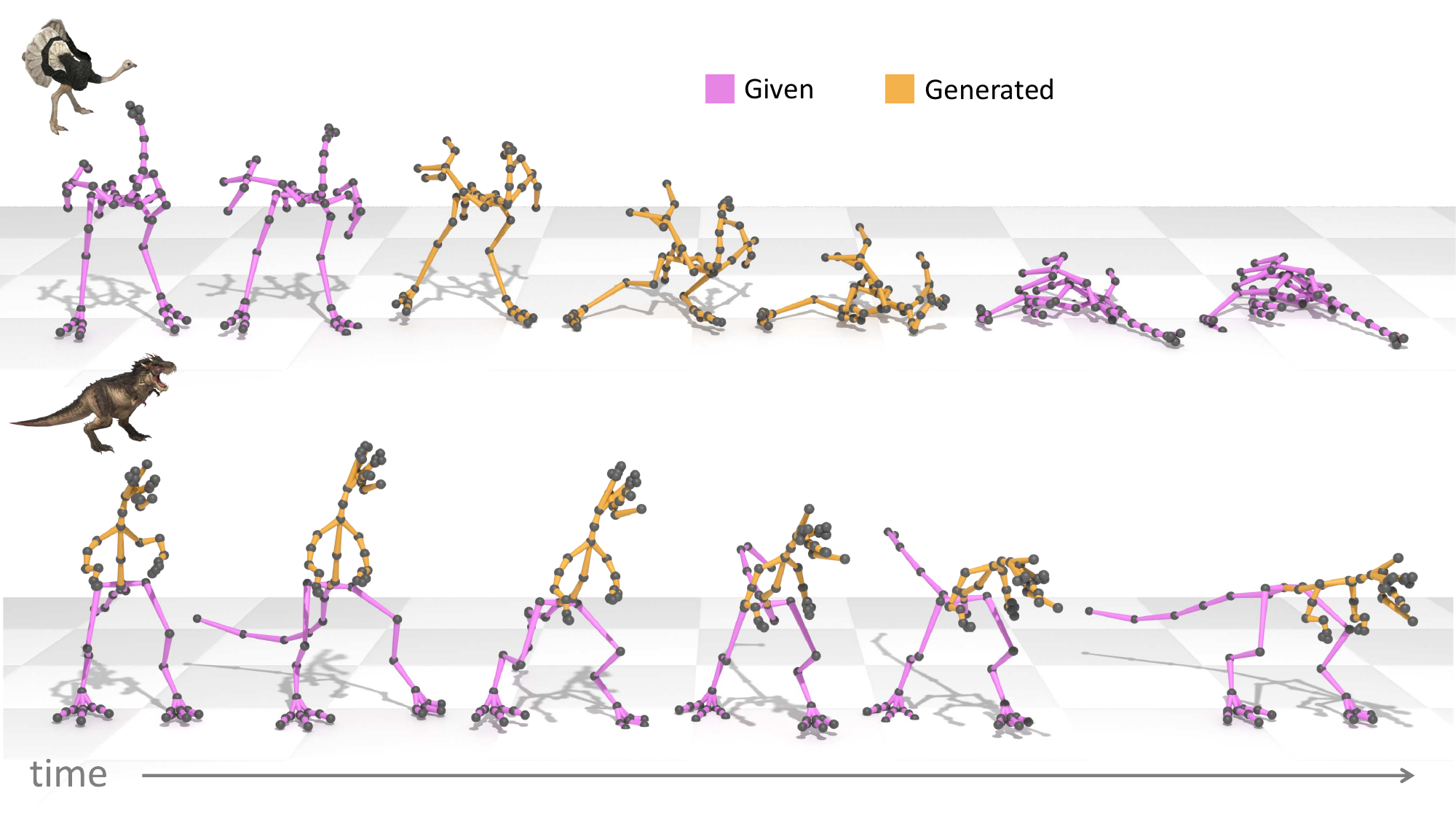}
    
    \caption{
        \textbf{Editing.} Top: In-betweening. Given motion prefix and suffix, \algoname can generate the middle frames. Bottom: Body part editing. Given the motion of the lower body, \algoname can generate its complement for the upper body.
        Both editing strategies produce smooth and natural transition between the given and the synthesized parts.
    }
    \label{fig:inpainting}
    \Description[]{}  %
\end{figure}

We demonstrate our method's versatility through two motion editing applications: \emph{in-betweening} for temporal manipulation and \emph{body-part editing}  for spatial modifications, both leveraging the same underlying approach.
For \emph{in-betweening}, the prefix and suffix of the motion are fixed, allowing the model to generate the middle.
For \emph{body-part editing}, we fix some of the joints and let the model generate the rest.
Given a fixed subset (temporal or spatial) of the motion sequence tokens, we override the denoised $\hat{x}_{0}$ at each sampling iteration with the fixed motion part. 
This approach ensures fidelity to the fixed For a clean sample $X_{0}$, either generated or given, and skeleton $\mss$, we extract DIFT features at diffusion step $t_{seg}\!=\!\!3$ and layer $l_{seg}\!=\!\!1$. The features are averaged along the joint dimension to produce a single feature vector per frame. We apply PCA for dimensionality reduction and then use K-means to cluster the frames into $k\!\!=\!\!3$ categories. Our results are visualized in \cref{fig:temoral_seg} and in the supp. video. while synthesizing the missing elements of the motion.
Our results, in \cref{fig:inpainting} and the supp. video, show a smooth and natural transition between the given and the synthesized parts, and demonstrate that our model successfully generalizes techniques previously limited to human skeletons \cite{tevet2023human} to accommodate diverse skeletal structures.

\section{Experiments}\label{sec:experiments}
\subsection{Dataset and Preprocessing} \label{sec:dataset}

The Truebones Zoo \cite{truebones} dataset comprises motion captures featuring 70 diverse skeletons, including mammals, birds, insects, dinosaurs, fish, and snakes. 
The number of motions per skeleton ranges from 3 to 40,
adding up to 1219 motions and 147,178 frames in total. 
The dataset includes variations in orientation, root definition, and scale. 
Additionally, the skeletons vary in joint order, naming conventions, and connectivity standards. 
To address these variations, we have performed comprehensive preprocessing of the data, including aligning all motions to the same orientation and average bone length, centering the first frame at the origin, and ensuring it is located on the ground.
This process is described in detail in \supp{sec:data}, and the processed data will be made available.

\paragraph{Skeletal Subsets} 
In addition to experimenting with the full dataset, 
we categorize the skeletons into four groups based on their motion dynamics and train \algoname on these subsets, alongside a model trained on the entire dataset.
The four skeletal categories are \emph{Quadrupeds}, \emph{Bipeds}, \emph{Flying}, and \emph{Insects}. 
These subsets allow us to constrain \crossgen to characters with similar behavior.
Our visualizations illustrate generations from models trained on the entire dataset or sub-datasets, depending on the context (\eg, \cref{fig:unseen}).

\subsection{Implementation details}
We use $T=100$ diffusion steps, $L=4$ \stt layers, and latent dimension $F=128$. 
We train the model using a single NVIDIA RTX A6000 GPU for 24 hours. Inference runs on an NVIDIA GeForce RTX 2080 Ti GPU. More implementation details can be found in \supp{sec:imp_detail}.

\subsection{Evaluation}
\paragraph{Benchmark}
To evaluate \algoname, we introduce a benchmark comprising 30 skeletons randomly selected from those with cumulative frame counts ranging between 600 and 1200. The benchmark includes 43\% Quadrupeds, 17\% Bipeds, 23\% Flying, and 17\% Insects,  reflecting the relative proportions of these categories in the dataset. 

\paragraph{Metrics} 
We report four metrics that measure different aspects of the generated motions, following \citet{raab2024single,li2022ganimator}. 
The metrics are calculated separately for each skeleton, and the mean and standard deviation across all tested skeletons are reported in the form \emph{$\text{mean}^{\pm\text{std}}$}. For each skeleton, we evaluate a number of samples proportional to its sample count in the dataset.
Let $M$,$G$ denote the group of ground truth (GT) and generated motions of the assessed skeleton, respectively. The metrics that we use are 
(a) \emph{coverage}, which is the rate of temporal windows in $M$, that are reproduced in $G$, (b) \emph{local diversity}, which is the average distance between windows in $G$ and their nearest neighbors in $M$, and (c) \emph{inter diversity}, the diversity between synthesized motions.
We define \emph{intra diversity} to be the diversity between sub-windows internal to a motion and define (d) \emph{intra diversity diff}, which is the difference between the intra diversity of $G$ and that of $M$.
Metrics (a) and (d) evaluate fidelity to the GT, while metrics (b) and (c) assess diversity. 
An ideal score features both high fidelity and high diversity.
High fidelity with low diversity suggests overfitting, while low fidelity with high diversity indicates divergence and noise.

\subsection{Baselines} 
To the best of our knowledge, no current works address such a diverse range of skeletal structures within a single model. Hence, we compare \algoname to adaptations of two baselines. The first is \emph{MDM} \cite{tevet2023human}, originally designed for a single humanoid skeleton. 
MDM uses per-frame embedding, so to match its representation format, we concatenate all joint features for each character, and pad them to a length of $J \times D$.
For fairness, we also concatenate the vectorized rest-pose embedding $\pp_{\mss}$ along the temporal axis as frame 0. Since MDM accepts textual conditions, we use the skeleton's name (e.g., Cat, Dragon) as the input text.
Additionally, since MDM's original configuration is designed for a dataset 14 times larger than ours \cite{guo2022generating}, we reduce its latent dimension size to mitigate overfit.

The second baseline is \emph{SinMDM} \cite{raab2024single}, designed to be trained on a \emph{single} motion sequence. We modify it to enable training on multiple sequences of the same character, resulting in a separate model for each skeleton. 

\subsection{Quantitative Results}
\begin{table}[b]
    \caption{\textbf{Comparison with baselines.} Our model clearly outperforms the baselines. \textbf{Bold} and \underline{underline} denote best and second best, respectively. $^*$ indicates the work was adapted to align with the terms of our experiment. 
    }
    \vspace{-5pt}
    \centering
    \resizebox{\columnwidth}{!}{
    \begin{tabular} {l | c c c c c}
         Model & Coverage $\uparrow$ & 
         \makecell{Local \\ Div.} $\uparrow$ & \makecell{Inter \\ Div. } $\uparrow$ & \makecell{Intra Div. \\ Diff.} $\downarrow$ & \makecell{\#Param. \\ (M)} $\downarrow$\\
        \midrule
        MDM$^*$ [\citeyear{tevet2023human}] & $71.3^{\pm31}$ & \underline{0.168}$^{\pm0.12}$ & $0.139^{\pm0.13}$ & $0.177^{\pm 0.08}$ & \underline{5.96}\\
        SinMDM$^*$ [\citeyear{raab2024single}]& \textbf{89.3}$^{\pm 15}$ & $0.080^{\pm 0.13}$ & \underline{0.280}$^{\pm 0.13}$ & \underline{0.144}$^{\pm 0.09}$ & 176.1 (5.87 $\times$ 30)\\
        \algoname (Ours) & \underline{80.5}$^{\pm 20}$ & \textbf{0.252}$^{\pm 0.14}$ & \textbf{0.312}$^{\pm 0.17}$ & \textbf{0.118}$^{\pm 0.07}$ & \textbf{2.28}\\
        \bottomrule
    \end{tabular}
    } %
    \label{tab:quant_comp}
\end{table}

\Cref{tab:quant_comp} shows a quantitative comparison of \algoname and the baselines.
\algoname outperforms MDM in all categories and SinMDM in all but coverage, which is expected since SinMDM is trained separately for each skeleton. 
Note the significant gap in diversity metrics, where the table shows \algoname generalizes well, while the others struggle to do so.
We also report the models' parameter count, showing ours uses fewer parameters, enabling lower computation and faster inference.
In \supp{sec:comp_subset}, we provide a comparison with the baselines on the data subsets, demonstrating our model's superiority on these as well.

\subsection{Qualitative Results}
Our supp. video reflects the quality of our results. 
It presents generated motions for various skeletons and comparisons to baselines.

\begin{figure}

    \centering
    
    \includegraphics[width=\columnwidth]{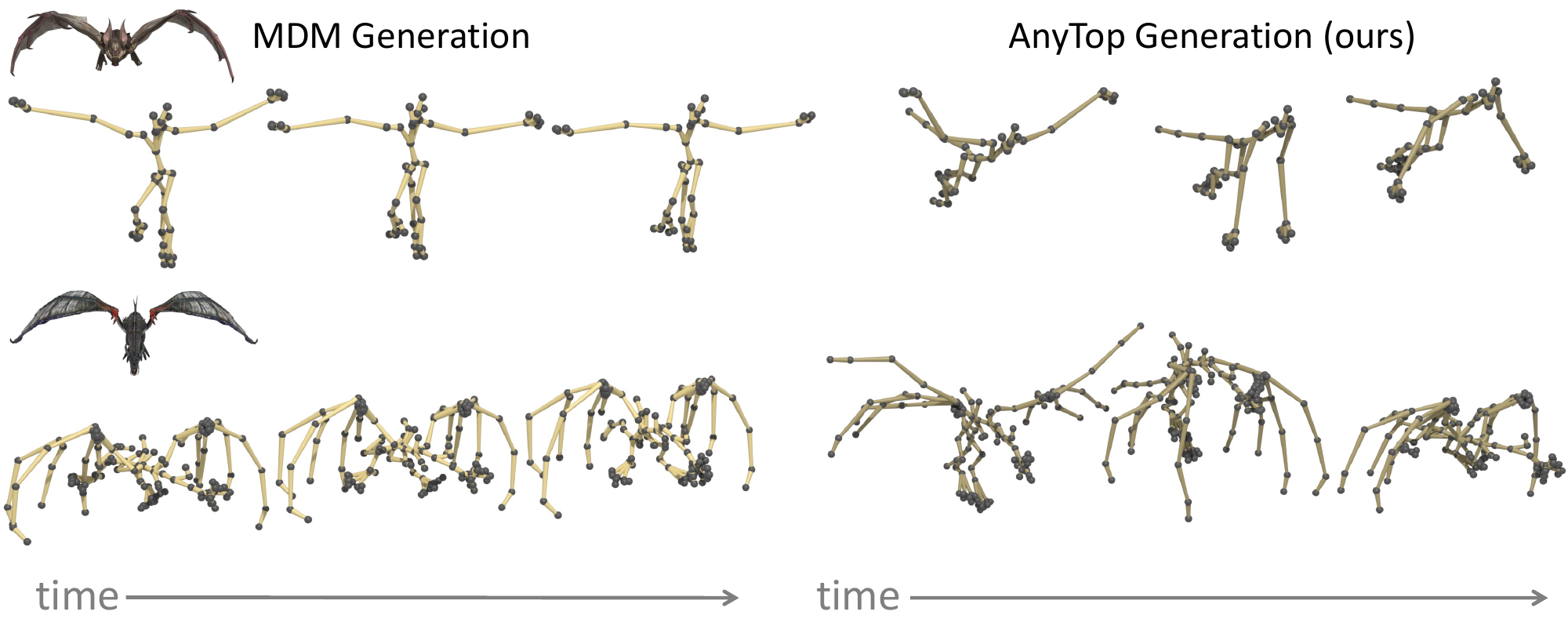}
    
    \caption{
        \textbf{Comparison with MDM Baseline.} \algoname (right) generates natural motions, while MDM (left) produces static, jittery motions.
    }
    \label{fig:baseline_comp}
    \Description[]{}  %
\end{figure}

In \cref{fig:baseline_comp}, we show that
\algoname produces natural and lively motions while MDM produces static, jittery motions. Moreover, MDM's results 
in our video 
show jittery transitions and unnatural poses.
\Cref{fig:in_gen} and our supp. video show that \algoname can generate novel poses by effectively combining joints from different ground truth poses.
In contrast, SinMDM is limited to temporal in-skeleton generalization and cannot handle spatial composition, due to its reliance on per-frame features.
Moreover, since SinMDM trains a separate model per skeleton, it cannot feature cross-skeleton or unseen-skeleton generalization.
As accurate foot contact is one of the
major factors of motion quality, we follow 
 \citet{li2022ganimator,raab2023modi} and use an IK post-process to ensure proper contact.

\paragraph{Unseen skeleton}
We present two unseen skeleton motions. One is a Komodo Dragon, generated by the \emph{Bipeds} model.
The second is a Cat, generated by a model trained on \emph{Quadrupeds}, excluding the cat.
\Cref{fig:unseen} and our supp. video demonstrate that \algoname generalizes well to unseen skeletons, while adapted MDM under the same settings generates static and jittery motions. 

\subsection{Ablation}

\begin{table}[b]
    \caption{\textbf{Ablation.} 
    Removing architectural choices leads to a degradation in \algoname's performance.
    }
    \vspace{-5pt}
    \centering
    \resizebox{\columnwidth}{!}{
    \begin{tabular} {l | c c c c}
         Model & Coverage $\uparrow$ & 
         \makecell{Local \\ Div.} $\uparrow$ & \makecell{Inter \\ Div. } $\uparrow$ & \makecell{Intra Div. \\ Diff.} $\downarrow$ \\
        \midrule
        \makecell[l]{w/o graph property \\ \quad embedding ($\dd, \Rcal$)} & 76.8$^{\pm{23}}$ & 0.249$^{\pm{0.14}}$ & \underline{0.303}$^{\pm{0.17}}$ & 0.127$^{\pm{0.11}}$ \\
        w/o rest-pose token & \underline{77.2}$^{\pm{25}}$& \underline{0.250}$^{\pm{0.14}}$ & 0.292$^{\pm{0.18}}$ & 0.130$^{\pm{0.09}}$\\
       \makecell[l]{w/o joint name \\ \quad embedding} & \textbf{82.3}$^{\pm{17}}$ & 0.218$^{\pm{0.12}}$ & 0.276$^{\pm{0.15}}$ & \textbf{0.113}$^{\pm{0.06}}$ \\
        
        \algoname (Ours) & \underline{80.5}$^{\pm 20}$ & \textbf{0.252}$^{\pm 0.14}$ & \textbf{0.312}$^{\pm 0.17}$ & \underline{0.118}$^{\pm 0.07}$ \\
        \bottomrule
    \end{tabular}
    } %
    \label{tab:ablation}
\end{table}

In \cref{tab:ablation}, we explore three key components of \algoname's architecture. 
First, the results confirm that without access to topological information, the model struggles to prioritize joints based on their hierarchical relations. Omitting the incorporation of $\dd$ and $\Rcal$ leads to degradation in all metrics.
Next, excluding the rest pose $\pp_{\mss}$ produces inferior results, reinforcing the idea that $\pp_{\mss}$ encodes vital information about joint offsets and bone lengths.
Lastly, we examine cross-skeletal \sr{textual} prior sharing via the addition of joint name embeddings. 
\sr{Incorporating joint names can be interpreted as balancing between in-skeleton and cross-skeleton generalization. Our ablation shows that excluding joint names (encouraging \ingen) increases coverage, whereas incorporating them (encouraging \crossgen) increases diversity.
By employing joint names, we enhance spatial correspondence, cross-generalization and unseen-generalization, acknowledging a minor decrease in coverage as an acceptable cost.}

\section{Conclusion, Limitations and Future Work}\label{sec:conclusion}

We have presented \algoname, a generative model that synthesizes diverse characters with distinct motion dynamics using a skeletal structure as input.
It uses a 
transformer-based denoising network, integrating graph information
at key points in the pipeline.
Our evaluation shows a highly informative latent space and notable generalization, even for characters with few or no training samples. 

One limitation of our method stems from imperfections in the input data. Despite our cleaning procedure, certain data artifacts remain unresolved\sr{, as detailed in \supp{sec:data}.} 
Another limitation is that our data augmentation process is computationally expensive with $O(J^2)$ complexity.

In the future, we plan to use \algoname for skeletal retargeting, multi-character interaction, editing, and various control modalities such as text-based and music-driven animation.
Another potential direction is editing
animations by simply modifying joint labels in the text descriptions.
Finally, future work could further explore DIFT features
in the motion domain.

\ifanonymous\else
    \begin{acks}
This research was supported in part by the Israel Science Foundation (grants no. 2492/20 and 3441/21), Len Blavatnik and the Blavatnik family foundation, and the Tel Aviv University Innovation Laboratories (TILabs).
\end{acks}

\fi

\bibliographystyle{ACM-Reference-Format}
\bibliography{bibiliography}

\ifappendix
    \appendix  %
    \newpage
    \section*{Appendix}  %
    This \ifappendix{Appendix }\else{Supplementary }\fi provides additional details to complement the information presented in the main paper. While the main paper is self-contained, the details provided here offer further insights and clarifications.

In \Cref{sec:imp_detail}, we provide implementation details of \algoname, and in \Cref{sec:data}, we elaborate on our preprocessing and augmentation pipelines. Finally, in \Cref{sec:comp_subset}, we present additional quantitative results beyond those in the main paper.

\sr{
\section{Related Work - Additional Details} \label{sec:rw_supp}
\begin{table*}
    \caption{
    \sr{
    \textbf{Methods categorized by supported skeletal variability.}     
    Each column features representative research works that support the type in its header, with full definitions of these categories provided in the main paper.
    }
    }
    \centering
    \resizebox{\textwidth}{!}{
    \begin{tabular} {l l l l l}
        Single & Isomorphic & Homeomorphic & Non-homeomorphic \\
        \midrule
        Text2Motion \cite{guo2022generating} & R$^2$ET \cite{zhang2023skinned} & SAN \cite{aberman2020skeleton}& MoMa \cite{martinelli2024moma} \\
        ACTOR \cite{petrovich2021actor} & \cite{Villegas2021ContactAwareRO} & SAME \cite{lee2023same} & WalkTheDog \cite{li2024walkthedog} \\
        GMD \cite{karunratanakul2023gmd} & SMPL \cite{loper2015smpl} & SMRNet \cite{zhang2024skinned} &  \\
        MAS \cite{kapon2023mas} & HUMOS \cite{tripathi2024humos} & CAR \cite{cao2024car} &  \\
        MoFusion \cite{dabral2022mofusion} & SCT \cite{jang2024geometry} & BMM \cite{studer2024factorized} &  \\
        MDM \cite{tevet2023human} & TEMOS \cite{petrovich2022temos} & DragPoser \cite{ponton2024dragposer} &  \\
        \cite{holden2016deep} & SMAL \cite{zuffi20173d} & \cite{zhang2024unified} &  \\
        MotionDiffuse \cite{zhang2024motiondiffuse} & OmniMotionGPT \cite{yang2023omnimotiongpt} & &  \\
        PriorMDM \cite{shafir2024human} & BARC \cite{rueegg2023barc}& &  \\
        MoDi \cite{raab2023modi} & & &  \\
        MoMask \cite{guo2024momask} & & & \\
        \bottomrule
    \end{tabular}
    } %
    \label{tab:rw_skel_type}
    \vspace{-5pt}
\end{table*}

\cref{tab:rw_skel_type} categorizes the methods listed in the main paper according to their supported skeletal variability.
}

\section{Implementation Details} \label{sec:imp_detail}
\begin{table}[b]
    \caption{\sr{\textbf{Ablation on the value of $d_{max}$.}}    
    }
    \centering
    \begin{tabular} {l | c c c c}
         $d_{max}$ & Coverage $\uparrow$ & 
         \makecell{Local \\ Div.} $\uparrow$ & \makecell{Inter \\ Div. } $\uparrow$ & \makecell{Intra Div. \\ Diff.} $\downarrow$ \\
        \midrule
        1 & 71.8 & \textbf{0.25} & 0.27 & \textbf{0.11} \\
        5 (Ours) & \textbf{80.5} & \textbf{0.25} & \textbf{0.31} & \textbf{0.11} \\
        10 & 79.3 & 0.23 & 0.29 & 0.12 \\
        \bottomrule
    \end{tabular}
    \label{tab:d_max}
    \vspace{-5pt}
\end{table}

The maximum topological distance we allow in $\dd$ is $d_{max}=5$, \sr{which is the closest to the dataset’s average kinematic chain length (4.65). In \cref{tab:d_max}, we show that indeed $d_{max}=5$ induces better scores.} Our \emph{\tempattn} is applied on temporal windows of length $W=31$, and for our model inputs, we allow a maximum number of joints $J=143$. \emph{During training}, we use cropped sequences of $N=40$ frames. To enable our model to handle higher frame positions and generate longer sequences, we incorporate positional encoding relative to the cropping index.
For training, we use a batch size of 16.

\section{Data} \label{sec:data}
\paragraph{Truebones Zoo dataset} In addition to the data misalignment issues discussed in the main paper, the dataset also contains vulnerabilities such as excessive dummy joints, qualitative artifacts like foot sliding and floating, and 20\% of the frames involve skeletons connected to the origin via an additional bone, resulting in artefacts such as walking or running in place. We address some of these issues as part of our data processing pipeline, which is detailed in the main paper and further extended in the following paragraph.

\paragraph{Data Preprocessing}
In this section, we provide further details on the preprocessing steps mentioned in the main paper, as well as describe additional refinements applied to the dataset. 
As part of the alignment process, we ensure that all skeletons are properly grounded. This is achieved by using the textual descriptions of the joints to identify the foot joints of each skeleton. Based on their height in the rest pose, we determine the ground height for each skeleton and subtract it from the corresponding root height in the motion data. For skeletons that do not interact with the ground, such as flying birds or swimming fish, the ground height is determined by the position of the lowest joint in the rest pose.
Another important preprocessing step is ensuring that the rest poses of all skeletons are natural. This is essential for two key reasons. 
First, many animals feature a similar span of rotation angles in organs that have similar functionality, \eg, the forearm. We would like this span of rotations to constitute a manifold representing multiple animals.
Once all rotation angles are defined relative to a character's rest-pose, we have a common representation basis, hence the desired manifold can be obtained.

Second, the rest pose is encoded as a single frame within the motion sequence. To maintain consistency with the other frames, which represent natural poses, the rest pose must also exhibit a natural configuration.
To accomplish this, we transform all motion rotations so that they are relative to a natural rest pose, which can either be provided as an additional motion capture (mocap) file or selected from the skeleton's idle motion.

In addition to the alignment procedure, we also extract relevant information from the skeletons and motion data. First, we use foot joints labels to generate foot-contact indicators for each frame, which are concatenated with the motion features. 

Next, we apply a normalization procedure. 
\sr{
Note that Truebones aggregates data from multiple sources, which leads to skeletons of inconsistent sizes, irrespective of their real-world proportions.}
This procedure involves computing the mean and standard deviation for each skeleton across \sr{all frames in all its motions} and using these statistics to normalize the motion data before training the model.

\sr{Despite our thorough cleaning procedure, several data artifacts remain unresolved in the ground-truth (GT) motions:
(1) inherent foot sliding effects (as in the walking chicken GT example in our supplementary video);
(2) artificial bones connecting characters to the ground;
(3) anatomical inconsistencies, such as right foreleg emerging from left shoulder and left foreleg from right shoulder;
(4) extraneous bones protruding from various body joints.}  

\paragraph{Input Preprocessing}
The input to our model is a skeleton $\mss = {\pp_\mss, \Rcal_\mss, \dd_\mss, \nn_\mss}$, derived from the raw rest pose of the character, represented as $(\mgg_\mss, O_\mss)$, along with the corresponding joint names. Both $(\mgg_\mss, O_\mss)$ and the joint names can be obtained from standard motion capture formats (\eg, bvh, fbx).

The skeletal features
$\mss = \{\pp_\mss, \Rcal_\mss, \dd_\mss, \nn_\mss\}$ are computed as follows: First, to compute $\pp_\mss$, we apply forward kinematics with zero rotations on $(\mgg_\mss, O_\mss)$ obtaining the global joint positions in the rest pose. These positions are then converted to root-relative coordinates. To align the rest pose with the format of individual frames in a motion sequence, we append to each root-relative position a 6D representation of zero rotation, zero velocity, and foot contact indicators. The topological conditions, $\Rcal_\mss$ and $\dd_\mss$, are derived through a traversal of the skeletal hierarchy $\mgg_\mss$. Finally, the joint names $\nn_\mss$ are extracted from the motion capture data and undergo text-preprocessing, which includes the removal of digits, symbols, irrelevant words, and redundant prefixes. Additionally, side indicators such as 'L/R' are replaced with 'Left/Right', non-English joint names are translated, and similar actions are standardized.

\paragraph{Data augmentation}
\emph{Skeletal Augmentation} exposes our model to a wider variety of skeletons, as described in the Method section of the main paper. Next, we further elaborate about this process. The first augmentation we apply is \emph{joint removal}, which randomly removes up to 30\% of the joints from the skeleton, where feet joints are never removed to maintain physical correctness. For efficiency, we exclude joints with multiple children from the removal procedure. We remove only those chains that terminate with end effectors, with the exception of feet. The second augmentation is \emph{joint addition}, which introduces a new joint at the midpoint of a randomly selected edge.
After removing or adding joints to the skeleton, we update $\Rcal_\mss$, $\dd_\mss$ and $\nn_\mss$ accordingly. Note that updating $\dd_\mss$ is computationally expensive with a complexity of $0(J^2)$, as it requires recomputing the path between each pair of joints in the DAG. 
\section{Comparison with Baselines on Subset Models} \label{sec:comp_subset}

\begin{table}[b]
    \caption{\textbf{Comparison on Data Subsets.} Quantitative results of \algoname trained on different data subsets, compared to the baselines trained under equivalent settings. $^*$ indicates the work has been adjusted to our experimental terms and $\dagger$ indicates that a specific skeleton (Scorpion) has been removed from the SinMDM evaluation set, as SinMDM fails to converge on this skeleton. This exclusion ensures that its impact does not skew the overall score.
    }
    \centering
    \resizebox{\columnwidth}{!}{
    \begin{tabular} {l l | c c c c}
         Subset & Model& Coverage $\uparrow$ & 
         \makecell{Local \\ Div.} $\uparrow$ & \makecell{Inter \\ Div. } $\uparrow$ & \makecell{Intra Div. \\ Diff.} $\downarrow$\\
        \midrule
        Quadrupeds & MDM$^*$ & 83.3$^{\pm{23}}$ & \underline{0.103}$^{\pm{0.14}}$ & 0.112$^{\pm{0.07}}$ & 0.160$^{\pm{0.03}}$\\
         &SinMDM$^*$ &  \textbf{94.0}$^{\pm{06}}$ & 0.050$^{\pm{0.04}}$ & \underline{0.230}$^{\pm{0.12}}$&\underline{0.151}$^{\pm{0.08}}$\\
        & \algoname & \underline{89.2}$^{\pm{09}}$ & \textbf{0.215}$^{\pm{0.08}}$ & \textbf{0.291}$^{\pm{0.17}}$ & \textbf{0.114}$^{\pm{0.06}}$\\
        
        \bottomrule

        Bipeds& MDM$^*$& 87.9$^{\pm{13}}$ & 0.034$^{\pm{0.01}}$ & 0.081$^{\pm{0.03}}$ & \underline{0.108}$^{\pm{0.05}}$\\
        & SinMDM$^*$& \textbf{95.0}$^{\pm{05}}$ & \underline{0.040}$^{\pm{0.02}}$ & \underline{0.251}$^{\pm{0.12}}$ & \textbf{0.090}$^{\pm{0.03}}$\\
        & \algoname & \underline{93.5}$^{\pm{05}}$ & \textbf{0.191}$^{\pm{0.09}}$ & \textbf{0.288}$^{\pm{0.19}}$ & 0.120$^{\pm{0.06}}$\\
        
        \bottomrule
        
        Flying& MDM$^*$& 63.7$^{\pm{31}}$ & \underline{0.219}$^{\pm{0.25}}$ & 0.193$^{\pm{0.18}}$ & \underline{0.154}$^{\pm{0.08}}$ \\
        & SinMDM$^*$& \textbf{78.9}$^{\pm{18}}$ & 0.071$^{\pm{0.04}}$ & \underline{0.320}$^{\pm{0.13}}$ & \textbf{0.095}$^{\pm{0.03}}$\\
        & \algoname & \underline{72.6}$^{\pm{18}}$ & \textbf{0.289}$^{\pm{0.13}}$ & \textbf{0.410}$^{\pm{0.19}}$ & 0.166$^{\pm{0.07}}$\\
        \bottomrule        
        Insects& MDM$^*$&88.4$^{\pm{07}}$&0.063$^{\pm{0.03}}$&0.185$^{\pm{0.10}}$&\textbf{0.117}$^{\pm{0.05}}$\\   
        & SinMDM $^*$ & 77.8$^{\pm{04}}$ & \textbf{0.235}$^{\pm{0.29}}$ & \textbf{0.419}$^{\pm{0.08}}$ & 0.152$^{\pm{0.05}}$\\
        & SinMDM $^{*\dagger}$ & \textbf{92.9}$^{\pm{03}}$ & 0.061$^{\pm{0.015}}$ & 0.348$^{\pm{0.10}}$ & 0.136$^{\pm{0.06}}$\\
   
        &\algoname & \underline{90.6}$^{\pm{09}}$ & \underline{0.189}$^{\pm{0.07}}$ & \underline{0.317}$^{\pm{0.117}}$ & \underline{0.127}$^{\pm{0.05}}$\\
        
        \bottomrule
    \end{tabular}
     } %
    \label{tab:sub_datasets}
\end{table}

We provide a quantitative evaluation of \algoname trained on the data subsets defined in the Experiments section of the main paper. To maintain fairness in comparison, we train the adapted MDM baseline separately for each subset. Since SinMDM is independently trained for each skeleton, no additional adjustments are needed. Each model is evaluated using the corresponding skeletons from our benchmark that match the relevant data subset. The results, shown in \cref{tab:sub_datasets}, indicate that \algoname achieves the optimal coverage-diversity tradeoff compared to all other baselines presented.

\fi
\end{document}